\documentclass[review]{elsarticle}

\usepackage{graphicx}
\usepackage{epstopdf}
\usepackage{subcaption}
\usepackage{color}
\usepackage{rotating}

\journal{Journal of \LaTeX\ Templates}

\begin{document}

\begin{frontmatter}

\title{Effect of the interplanetary magnetic field orientation and intensity in the mass and energy deposition on the Hermean surface}


\author[label1,label2]{J. Varela}
\ead{deviriam@gmail.com (telf: 0033782822476)}

\author[label2]{F. Pantellini}
\author[label2]{M. Moncuquet}

\address[label1]{AIM, CEA/CNRS/University of Paris 7, CEA-Saclay, 91191 Gif-sur-Yvette, France}
\address[label2]{LESIA, Observatoire de Paris, CNRS, UPMC, Universite Paris-Diderot, 5
place Jules Janssen, 92195 Meudon, France}

\begin{abstract}

The aim of the present study is to simulate the interaction between the solar wind and the Hermean magnetosphere. We use the MHD code PLUTO in spherical coordinates with an axisymmetric multipolar expansion of the Hermean magnetic field, to perform a set of simulations with different interplanetary magnetic field orientations and intensities. We fix the hydrodynamic parameters of the solar wind to study the distortions driven by the interplanetary magnetic field in the topology of the Hermean magnetosphere, leading to variations of the mass and energy deposition distributions, the integrated mass deposition, the oval aperture, the area covered by open magnetic field lines and the regions of efficient particle sputtering on the planet surface. The simulations show a correlation between the reconnection regions and the local maxima of plasma inflow and energy deposition on the planet surface.

\end{abstract}


\end{frontmatter}


\section{Introduction}
\label{Introduction}

First measurements of MARINER 10 magnetometer identified a dipolar moment of the Hermean magnetic field of $195nT*R^{3}_{M}$ ($R_{M}$ Mercury's radius) \citep{2008Sci...321...85S}, as well as a relative small proportion between the intensity of the interplanetary magnetic field (IMF) and the Hermean magnetic field $\alpha = B_{sw}//B_{M}$ \citep{1974Sci...185..131N}. Further refined observations during thousands of MESSENGER orbits lead to model the Hermean magnetic field as a multipolar expansion \citep{2012JGRE..117.0L12A}. The IMF intensity oscillates between a range of few nT and almost 60 nT during coronal mass ejections (CME) \citep{2009JGRA..11410101B,2011PandSS...59.2066B}, together with IMF orientations that can depart from the Parker Spiral \citep{1958ApJ...128..664P}, which entails a rich variety of Hermean magnetosphere configurations \citep{1979JGR....84.2076S,2007SSRv..132..529F}.

Observational studies measured the temporal and latitudinal variability of charged particles precipitation on the Hermean surface \citep{2012AGUFM.P32B..08B,2012LPI....43.1646D}, the magnetosheath depletion \citep{1982GeoRL...9..921B,2013JGRA..118.7181G}, the plasma injection and transport in the inner magnetosphere, as well as the evolution of the Hermean magnetic field topology for different solar wind (SW) and IMF configurations \citep{2013AGUFMSM24A..03D}. 

The interaction of the SW and the Hermean magnetosphere was studied using different numerical frameworks, for example, single fluid MHD models \citep{2008Icar..195....1K,2008JGRA..113.9223K,2015JGRA..120.4763J}, multifluid MHD models \citep{Muller2011946,Muller2012666} and hybrid models \citep{2012JGRA..11710228R,2010Icar..209...46W}. The simulations results show an enhancement or a weakening of the Hermean magnetic field according to the IMF orientation \citep{1979JGR....84.2076S,2000Icar..143..397K,2009Sci...324..606S}, that imply distortions in the Hermean magnetosphere topology and in the plasma flows towards the planet surface \citep{2003Icar..166..229M,2003GeoRL..30.1877K,2004AdSpR..33.2176K,2007AGUFMSM53C1412T,2010Icar..209...11T}. Previous studies are limited to specific SW and IMF configurations, so a parametric analysis is required to fully assess the effect of the IMF orientation and intensity on the Hermean magnetosphere.

The aim of present study is to analyze the effect of the IMF orientation and intensity on the Hermean magnetosphere topology. We identify the role of the reconnection regions in the SW injection into the inner magnetosphere, the mass and energy deposition, the oval angle aperture and the regions of efficient particle sputtering on the planet surface. We perform a set of simulations which IMF is oriented in three possible directions: a Sun-Mercury direction, a direction parallel to the planet magnetic axis and a direction parallel to the planet orbit plane (perpendicular to the other orientations). In addition, the IMF intensity can take the values 10, 20 and 30 nT. The hydrodynamic SW parameters in the simulations are obtained from the numerical model ENLIL + GONG WSA + Cone SWRC during the MESSSENGER orbit of the date 2011/10/10 \citep{2013JGRA..118...45B}. We identify the reference case of the analysis with a configuration without IMF.

Present study is a complement and extension of past author's communications devoted to analyze the correlation between reconnection regions in the Hermean magnetosphere and local inflow maxima on the planet surface, emulating the IMF and SW conditions during several MESSENGER orbits \citep{2015PandSS..119..264V}. We also studied the properties of plasma streams originated in the magnetosheath \citep{2016PandSS..120...78V} and the effect of the SW hydrodynamic parameters on the plasma precipitation towards the Hermean surface \citep{Varela201646}.

We use the MHD version of the single fluid code PLUTO in spherical 3D coordinates \citep{2007ApJS..170..228M}. The Northward displacement of the Hermean magnetic field is represented by an axisymmetric multipolar expansion \citep{2011Sci...333.1859A}. 

This paper is structured as follows. Section II, a description of the numerical model, boundary and initial conditions. Section III, a parametric study for several IMF configurations, to analyze the effect of the IMF orientation and intensity on the Hermean magnetic field topology and plasma flows towards the planet surface. Section IV, conclusions and discussion.

\section{Numerical model}
\label{Model}

We use the MHD version of the open source code PLUTO in spherical coordinates to simulate a single fluid polytropic plasma in the non resistive and inviscid limit \citep{2007ApJS..170..228M}.

The conservative form of the equations are integrated using a Harten, Lax, Van Leer approximate Riemann solver (hll) associated with a diffusive limiter (minmod). The divergence of the magnetic field is ensured by a mixed hyperbolic/parabolic divergence cleaning technique \citep{2002JCoPh.175..645D}.

The models grid points are: $196$ radial points, $48$ in the polar angle $\theta$ and $96$ in the azimuthal angle $\phi$ (the grid poles correspond to the magnetic poles). The numerical magnetic Reynolds number of the simulations due to the grid resolution is $R_{m}= V L/\eta \approx 1350$, with $V = 10^{5}$ m/s and $L = 2.44 \cdot 10^{6}$ m the characteristic velocity and length of the model, and $\eta \approx 1.81 \cdot 10^{8}$ m$^{2}$/s the numerical magnetic diffusivity of the code. The numerical magnetic diffusivity was evaluated in dedicated numerical experiments using a model with the same grid resolution but a simpler setup.

We use a multipolar expansion of the planet magnetic field assuming an axisymmetric model. The magnetic potential $\Psi$ is expanded in dipolar, quadrupolar, octupolar and hexadecapolar terms \citep{2011Sci...333.1859A}:

\begin{equation} \label{eq:1}
\Psi (r,\theta) = R_{M}\sum^{4}_{l=1} (\frac{R_{M}}{r})^{l+1} g_{l0} P_{l}(cos\theta)  
\end{equation}

The current free magnetic field is $B_{M} = -\nabla \Psi $. $r$ is the distance to the planet center, $\theta$ the polar angle and $P_{l}(x)$ the Legendre polynomials. The numerical coefficients $g_{l0}$ taken from Anderson et al. 2012 are summarized in Table 1.

\begin{table}[h]
\centering
\begin{tabular}{c | c c c c}
coeff & $g_{01}$(nT) & $g_{02}/g_{01}$ & $g_{03}/g_{01}$ & $g_{04}/g_{01}$  \\ \hline
 & $-182$ & $0.4096$ & $0.1265$ & $0.0301$ \\
\end{tabular}
\caption{Multipolar coefficients $g_{l0}$ for Mercury's internal field.}
\end{table}

The simulation frame is such that the z-axis is given by the planetary magnetic axis pointing to the magnetic North pole and the Sun is located in the XZ plane with $x_{sun} > 0$. y-axis completes the right-handed system. 

The simulation domain is confined within two spherical shells centred on the planet, representing the inner and outer boundaries of the system. The shells are at $0.6 R_{M}$ and $12 R_{M}$. Between the inner shell and the planet surface (at radius unity in the domain) there is a "soft coupling region" where special conditions apply. The outer boundary is divided in two regions, upstream part where solar wind parameters are fixed and downstream part where we consider null derivative condition $\frac{\partial}{\partial r} = 0$ for all fields ($\vec{\nabla} \cdot \vec{B} =0$ condition is not violated because at $12 R_{M}$ the IMF is dominant and constant). At the inner boundary the value of the intrinsic magnetic field of Mercury is specified. In the soft coupling region the velocity is smoothly reduced to zero when approaching the inner boundary, setting magnetic and velocity fields parallel. Density is adjusted to keep Alfven velocity constant $\mathrm{v}_{A} = B / \sqrt{\mu_{0}\rho} = 25$ km/s with $\rho = nm_{p}$ the mass density, $n$ the particle number, $m_{p}$ the proton mass and $\mu_{0}$ the vacuum magnetic permeability. In the initial conditions we define a paraboloid on the night side with the vertex at the center of the planet, defined as $r < 1.5 - 4sin(\theta)cos(\phi) / (sin^{2}(\theta)sin^{2}(\phi)+cos^{2}(\theta))$, where the velocity is null and the density is two orders smaller compared to the solar wind. IMF is cut off at $2 R_{M}$.

The solar wind parameters of the simulations are summarized in Table 2. We assume fully ionized proton-electron plasma. The sound speed is defined as $\mathrm{v}_{s} = \sqrt {\gamma p/\rho} $ (with $p$ the total electron + proton pressure), the sonic Mach number as $M_{s} = \mathrm{v}/\mathrm{v}_{s}$ with $\mathrm{v}$ the velocity. The IMF in the reference case is null and the solar wind velocity is aligned with the Sun-Mercury direction for simplicity in all the simulations.

\begin{table}[h]
\centering
\begin{tabular}{c | c c c c c c c c}
Date & $n$ (cm$^{-3}$) & $T$ (K) & $\mathrm{v}$ (km/s) & $M_{s}$  \\ \hline
2011/10/10 & $60$ & $58000$ & $250$ & $6.25$ \\
\end{tabular}
\caption{Reference simulation parameters}
\end{table}

Any analysis performed in present communication is done after the model evolution reaches a robust steady state. SW properties are kept constant during the simulations and the study doesn't describe any dynamic event of the Hermean magnetosphere. The transition from the initial conditions to the steady state only shows the numerical adjustment of the model without any physical valuable information, so it is not included in the text. The simulations require $t \approx 200$ s of evolution to reach the steady state (with "t" the non normalize time of the simulation).

\section{Parametric study}
\label{Parametric study}

In this section we study the effect of the IMF orientation and intensity on the Hermean magnetic field topology together with the flows toward the planet surface. We analyze the properties of the plasma stream that binds magnetosheath and Mercury's surface.

In the following we identify the Mercury-Sun orientation as Bx simulations, the Sun-Mercury orientation as Bxneg simulations, the Northward orientation respect to Mercury's magnetic axis as Bz simulations, the Southward orientation as Bzneg simulations, the orientation perpendicular to previous two cases on the planet orbit plane as By and Byneg simulations. The IMF intensity of the model is denoted by a number attached to the orientation label, for example the simulation Bz identifies a Northward IMF orientation of module 10 nT. Figure 1 shows a 3D view of the system. The red lines identify the Hermean magnetic field lines, the green lines the SW stream lines and the color scale the density distribution in XZ plane to show the region of the bow shock (BS).

\begin{figure}[h]
\centering
\includegraphics[width=0.5\textwidth]{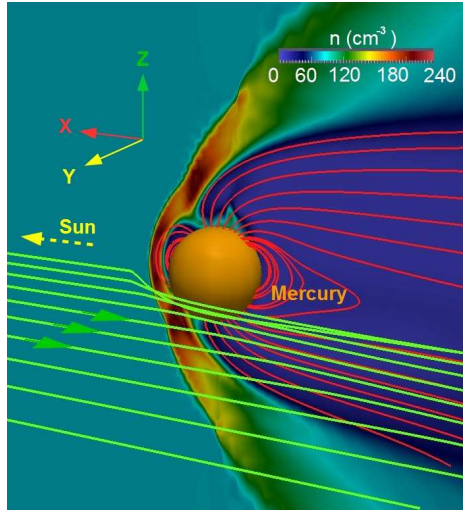}
\caption{3D view of the system. Hermean magnetic field lines (red lines), SW stream lines (green lines) and density distribution in XZ plane (color distribution).}
\end{figure}

\subsection{Hermean magnetic field topology}

First, we analyze the effect of the IMF orientation on the Hermean magnetic field topology and global structures of the magnetosphere. Figure 2 shows polar plots of the density distribution including the magnetic field lines of the planet (red lines) and SW stream lines (green lines).

\begin{figure}[h]
\centering
\includegraphics[width=1.0\textwidth]{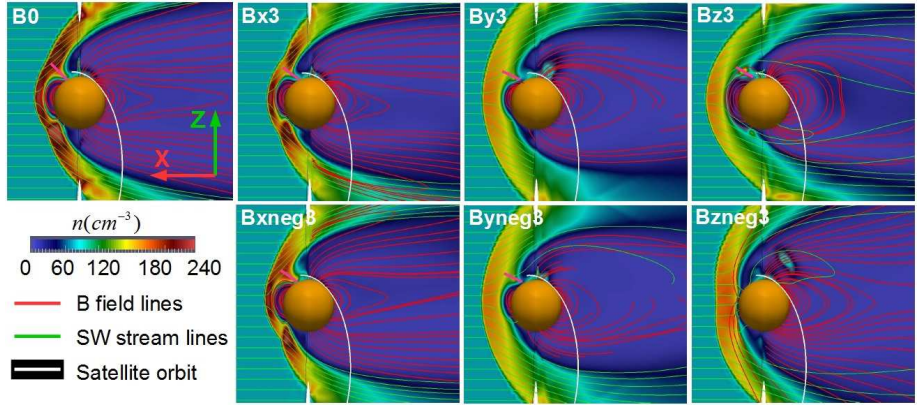}
\caption{Polar plot of the density distribution (displaced $0.1R_{M}$ in $Y$ direction) for the reference case and the simulations Bx3, Bxneg3, By3, Byneg3, Bz3 and Bzneg3. The red lines identify the Hermean magnetic field lines, the green lines the SW stream lines and the white line the satellite reference orbit. The pink line indicates the strip of the model plotted on Figure 8.}
\end{figure}

The SW reaches the planet surface at the equator only in Bzneg3 simulation, no close magnetic field lines on the day side, due to the strong reconnection of IMF and planetary magnetic field. The magnetic diffusion of the numerical model is several orders of magnitude larger than the magnetic diffusion of the real plasma, so the reconnection between the IMF and the Hermean magnetosphere in the simulations is instantaneous (no magnetic pile-up on planet dayside) and more intense (enhanced erosion of the Hermean magnetic field). Despite this fact, the model reproduces the main features of the plasma injection into the inner magnetosphere through the reconnection regions and the magnetosheath plasma depletion. A detailed discussion of the plasma injection and magnetosheath plasma depletion is provided in the description of figures 3 and 8, as well as in the conclusions section. For further information see reference \citep{Varela201646}. All the other simulations show close lines on the day side. The stand off distance (distance from the magnetopause to the planet surface at the equatorial plane) is almost the same in Bx3-Bxneg3 simulations and the reference case, slightly smaller in By3-Byneg3 simulations and located further away in Bz3 simulation. The magnetotail is curved to the South (North) if the IMF is oriented towards Bx (Bxneg) direction. The magnetotail shows an East-West asymmetry for By and Byneg orientations. The magnetotail is wider for Bz IMF orientations although it is smaller and the reconnection X point is located closer to the planet surface for Bzneg IMF orientations.

Figure 3 shows the Hermean magnetic field lines (magnetic field intensity imprinted on the field lines), SW stream lines (green lines), magnetic field intensity at the frontal plane $X = 0.3R_{M}$ and inflow/outflow (blue/red) regions on the planet surface (same simulations than Fig. 2). The regions of SW injection into the inner magnetosphere (reconnection regions are identified as blue colors at the frontal plane) are connected to the stream lines of the plasma column toward the planet surface. The South Hemisphere is more exposed to the SW because the magnetic field of Mercury is displaced to the North along the planet rotation axis, feature reproduced by the multipolar configuration of the Hermean magnetic field of the model. 

\begin{figure}[h]
\centering
\includegraphics[width=1.0\textwidth]{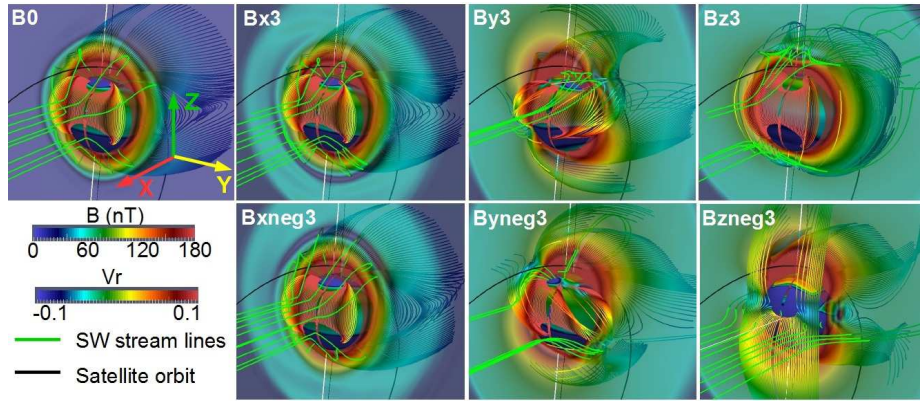}
\caption{Hermean magnetic field lines with the intensity imprinted on the field lines by a color scale (same simulations than Fig. 2). Magnetic field intensity at the frontal plane $X = 0.3R_{M}$. SW stream lines (green). Inflow/outflow regions on the planet surface (blue/red). Satellite's trajectory (black line).}
\end{figure}

The reference model, no IMF, lacks regions of reconnection and plasma injection into the inner magnetosphere. The Hermean magnetosphere topology is not distorted by the IMF so no regions of extra plasma precipitation on the planet surface are observed. The inflow regions on the planet surface are located nearby the poles, caused by plasma falling through the cusp. By contrast, there is a wide reconnection region nearby the equator in Bzneg3 simulation, reason why no close magnetic lines are observed on the day side of the planet. An inflow region covers main part of the Hermean surface, connected with SW stream lines (green lines), indicating that the SW precipitates directly on the planet surface. Bx3 (Bxneg3) simulation shows a wide reconnection region in the South (North) of the magnetosphere. A plasma column falls toward the planet surface, linked to wider inflow regions near the South (North) pole in Bx3 (Bxneg3) simulation compared to the reference case. By3 (Byneg3) simulation shows an East-West asymmetry of the Hermean magnetic field, observed too in the location of the reconnection regions, closer to the North pole in the East (West) side of the planet and to the South pole in the West (East) side, leading to a East-West displacement of the inflow local maxima on the planet surface compared to the reference case. In Bz3 simulation the reconnection regions are located closer to the poles and the local inflow maxima is smaller, particularly at the North Hemisphere, because the Hermean magnetic field is enhanced at the equatorial region and weakened near the poles. 

Figure 4 shows the magnetic field module and components at planes rotated $0^{o}$ (A-D), $30^{o}$ (E-H) and $60^{o}$ (I-M) with regard to the equatorial plane at the North Hemisphere (same simulations than Fig. 2).

\begin{figure}[h]
\centering
\includegraphics[width=0.8\textwidth]{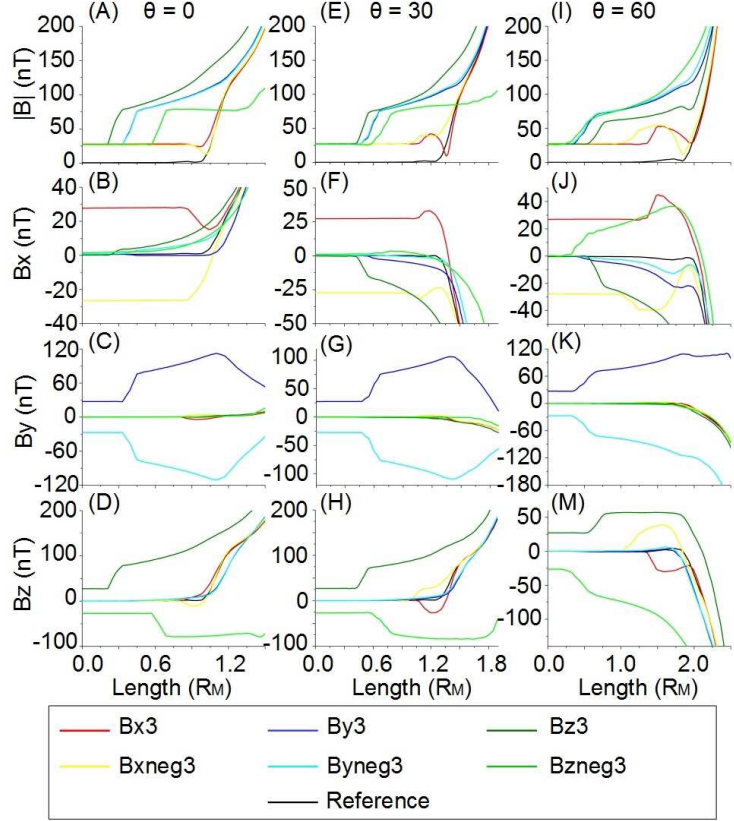}
\caption{Magnetic field module and components at planes rotated $0^{o}$ (A-D), $30^{o}$ (E-H) and $60^{o}$ (I-M) with regard to the equatorial plane at the North Hemisphere (same simulations than Fig. 2).}
\end{figure}

The profiles of the magnetic field module in the reference case, simulations Bx3 and Bxneg3 are very similar, particularly nearby the equator, pointing out that Bx and Bxneg IMF orientations only modify the intensity of the Hermean magnetic field close to the poles. In contrast, if we compared Bx and Bxneg IMF orientations to the reference case in the magnetosheath region, the magnetic field topology is different showing opposite rotations of the magnetic field components. The profiles of the magnetic field module in By3 and Byneg3 simulations are very similar between them but different compared to the reference case. The Hermean magnetic field topology is distorted mainly through $B_{y}$ component, because the other components show similar profiles than the reference case, leading to a West-East asymmetry of the magnetosphere. For the Bz3 (Bzneg3) simulation, the Hermean magnetic field module increases (decreases) at the equator and increases (decreases) at the poles. Hermean magnetic field topology is strongly distorted by Bz (Northward) and Bzneg (Southward) IMF orientations, driving different rotations of the magnetic field components compared to the reference case.

Figure 5 shows the BS (SI) and magnetopause (MI) location at planes rotated $0^{o}$ (A-D), $45^{o}$ (B-E) and $90^{o}$ (C-F) with regard to the equatorial plane at the North Hemisphere. If we compare the SI and MI profiles of a Bx-Bxneg orientation to the reference case, if the IMF intensity increases, the magnetosheath compression is stronger at low latitudes and weaker at high latitudes. For a By-Byneg orientation, the magnetosheath is less compressed if the IMF intensity increases and the magnetopause is located closer to the planet. For a Bz orientation, the magnetosheath compression decreases if the IMF intensity increases, located further away from the planet at low latitudes and closer at high latitudes. For a Bzneg orientation, the magnetopause reaches the planet surface at the equator if the IMF intensity is $20$ nT, showing a decompression of the magnetosheath. The magnetopause is closer to the planet at low latitudes if the IMF intensity increases although it is located further away at high latitudes.

\begin{figure}[h]
\centering
\includegraphics[width=0.8\textwidth]{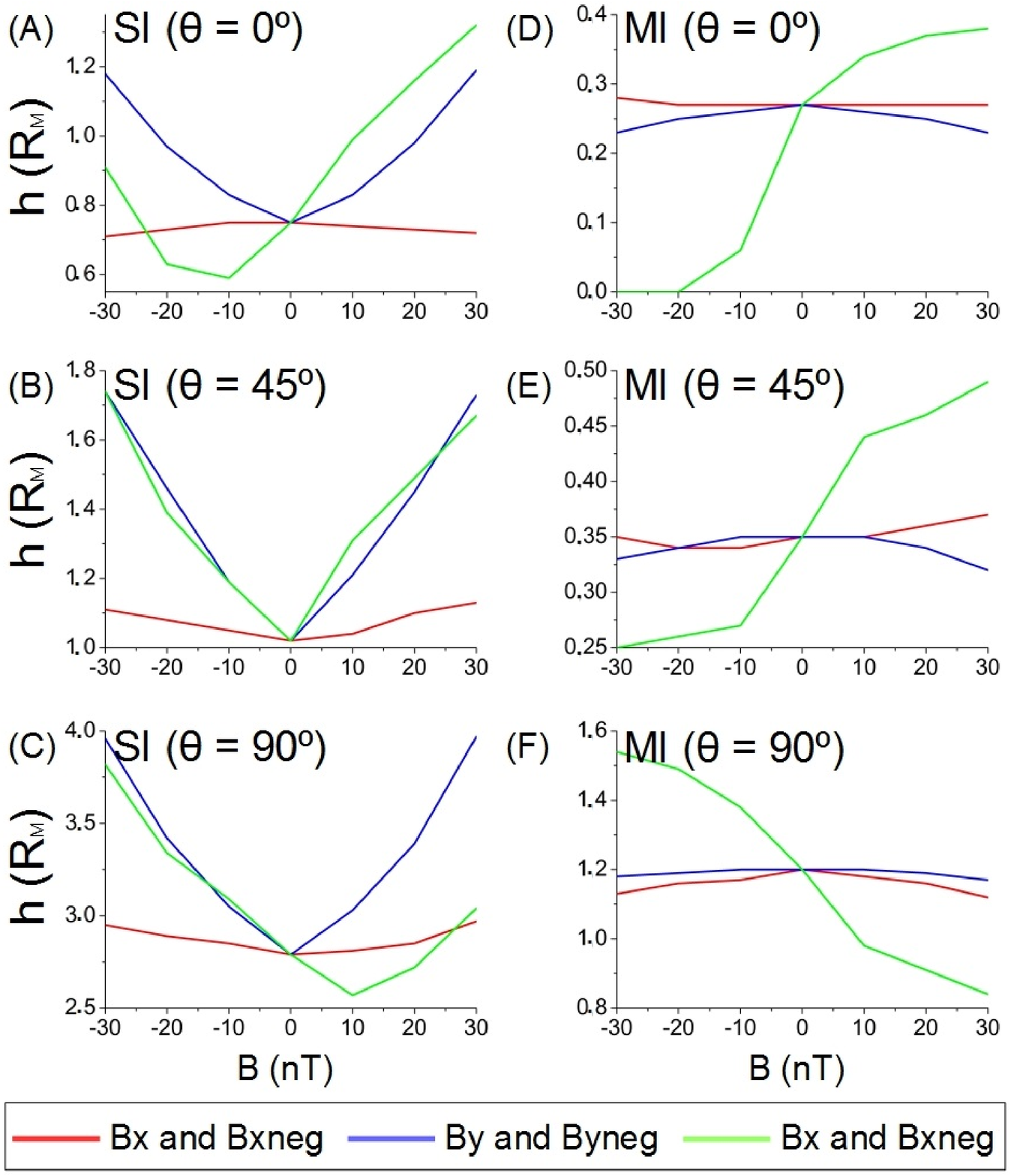}
\caption{Location of the BS (SI) and the magnetopause (MI) at planes rotated $0^{o}$ (A-D), $45^{o}$ (B-E) and $90^{o}$ (C-F) with regard to the equatorial plane at the North Hemisphere for different IMF orientations and intensities. $h(R_{M})$ is the height above the planet surface.}
\end{figure}

Figure 6 shows the area of the planet surface covered by open magnetic field lines. This area increases with the IMF intensity for Bx and Bxneg IMF orientations. The reconnection region in the South (North) of the magnetosphere for a Bx (Bxneg) IMF orientation leads to a larger exposition of the South (North) Hemisphere as the IMF intensity increases. By-Byneg orientations show a similar increase of the exposed area as the IMF intensity is enhanced. The exposed area is almost 3 times larger at the South Hemisphere and this ratio remains unchanged if the IMF intensity increases. For a Bz (Bzneg) orientation, the exposed area decreases (increases) with the IMF intensity. The ratio of exposed area between Hemispheres decreases (increases) if we compare the simulation Bzneg3 (Bz3) to the reference case, $1/3$ ($3$) times versus $2.5$ times.

\begin{figure}[h]
\centering
\includegraphics[width=0.8\textwidth]{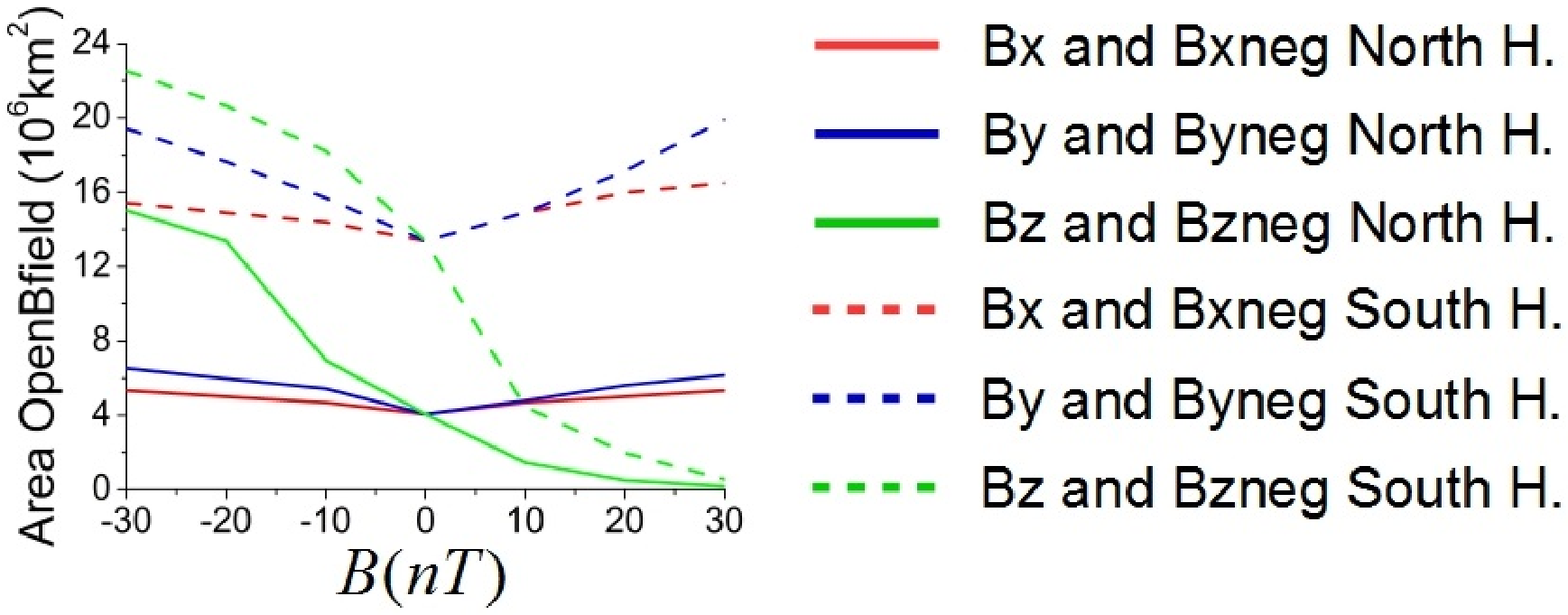}
\caption{Area of planet surface with open magnetic field lines for different IMF orientations and intensities.}
\end{figure}

Figure 7 shows the oval aperture angle at the North (A) and South Hemispheres (B). The oval aperture angle at the North Hemisphere increases with the IMF intensity for a Bx-Bxneg orientation, around $5^{o}$ if we compare Bx3 and Bxneg3 simulations to the reference case. A By-Byneg orientation shows different oval angles in the West and in the East of the Hemispheres due to the West-East asymmetry of the magnetosphere. There is a difference of $12^{o}$ in the West (East) and $5^{o}$ in the East (West) if we compare By3 (Byneg3) simulation to the reference case. The oval angle for a Bz-Bzneg orientation increases (decreases) with the IMF intensity on the day (night) side of the planet. All the day side is covered by open field lines in Bzneg3 simulation, although the oval angle decreases in $9^{o}$ in Bz3 simulation. On the night side, the oval angle is $15^{o}$ in Bzneg3 simulation and $9^{o}$ in Bz3 case. The oval angle profiles show the same trends for different IMF intensities and orientations at the South and North Hemisphere, but the angles are larger at the South Hemisphere due to the Northward displacement of the Hermean magnetic field (oval angle of $50^{o}$ in the reference case). The oval angles in Bx3 and Bxneg3 simulations are $5^{o}$ larger compared to the reference case. The oval angle in By3 (Byneg3) simulation is $20^{o}$ larger in the West (East) side and $1^{o}$ in the East (West) side. Bz3 simulation shows an oval angle $36^{o}$ smaller on the day and night side compared to the reference case. The oval angle in Bzneg2 simulation covers all the day side although it is $22^{o}$ smaller than the reference case on the night side.

\begin{figure}[h]
\centering
\includegraphics[width=0.8\textwidth]{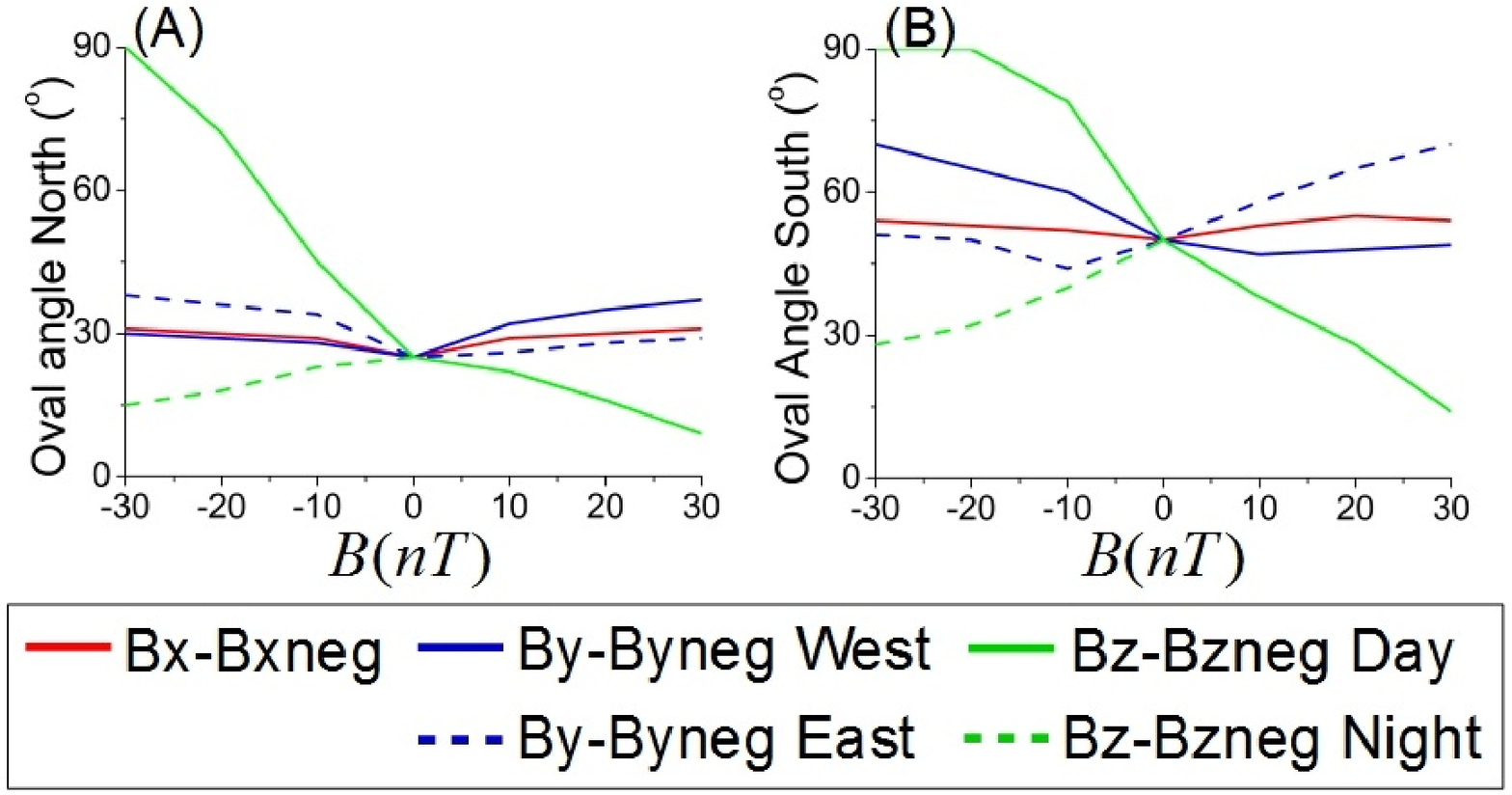}
\caption{Oval angle at the North (A) and South Hemispheres (B) for different IMF orientations and intensities.}
\end{figure}

\subsection{Plasma flows toward the planet surface}

After analyzing the perturbations driven in the Hermean magnetosphere topology by different IMF orientations and intensities, in this section we describe the ulterior consequences in the flows towards Mercury's surface. Figure 8 shows the evolution of the density, temperature, magnetic and velocity field (module and components) along the plasma stream, from the magnetosheath (on the left side of the graphs) to the planet surface (on the right side of the graphs). We perform the study for the same simulations than Fig. 2 except Bzneg3 simulation, because no plasma stream is observed in this model.

\begin{figure}[h]
\centering
\includegraphics[width=0.6\textwidth]{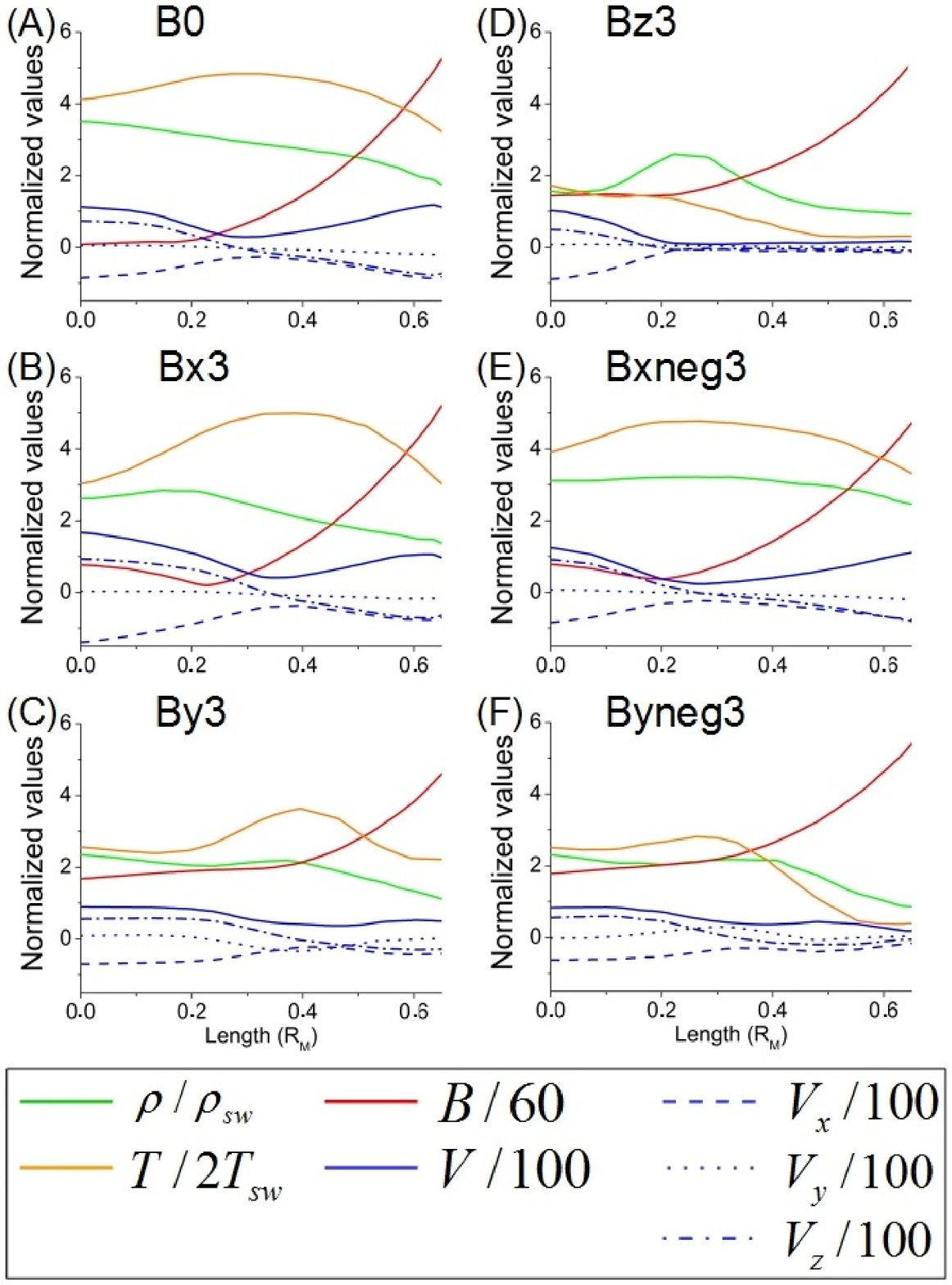}
\caption{Density, temperature, magnetic and velocity field module and components) along the plasma stream (see figure 2, pink lines).}
\end{figure}

The plasma stream originates from the magnetosheath near the reconnection region, identified on the graphs as a drop or a flattening of the magnetic field module profile (according to the plotted strip is closer or further away to the reconnection region). Nearby the reconnection region, the plasma in the magnetosheath is more dense, it is decelerated and heated before falling via the open magnetic field lines toward the planet surface. During the downfall the plasma is less dense, it is cooled and accelerated ($V_{z}$ component is as large as $V_{x}$). The simulations with the fastest flows toward the planet surface are the reference, Bx3 and Bxneg3 models although the most dense flows are observed in Bxneg3 simulation and reference case. These results suggest that the reference case should be the simulation with the largest mass deposition at the North Hemisphere, but first it is mandatory to analyze the size of the plasma deposition regions.

Figure 9 shows the radial velocity distribution and the area covered by open magnetic field lines on the Hermean surface. If we compare Bx3 (Bxneg3) simulation to the reference case, the size of the inflow area increases at the South (North) Hemisphere and decreases at the North (South) Hemisphere due to the presence of the reconnection region. The planet surface covered by open magnetic field lines at the North Hemisphere is slightly wider (narrower) in Bx3 (Bxneg3) simulation but narrower (wider) at the South Hemisphere. In By3 (Byneg3) simulation the inflow regions are moved to the East (West) at the North Hemisphere and to the West (East) at the South Hemisphere, as well as the surfaces covered by open magnetic field lines, wider in the West (East) of the North Hemisphere and in the East (West) of the South Hemisphere. In Bz3 simulation, inflow and open magnetic field lines regions are localized nearby the poles, although in Bzneg3 simulation the largest inflow region is observed on the day side at the planet equator, slightly moved to the North Hemisphere, with open magnetic field lines covering all the planet day side. In summary, the size of the deposition region at the North Hemisphere is wilder in Bxneg3 simulation compared to the reference case, so the mass deposition it is bigger too.

\begin{figure}[h]
\centering
\includegraphics[width=0.8\textwidth]{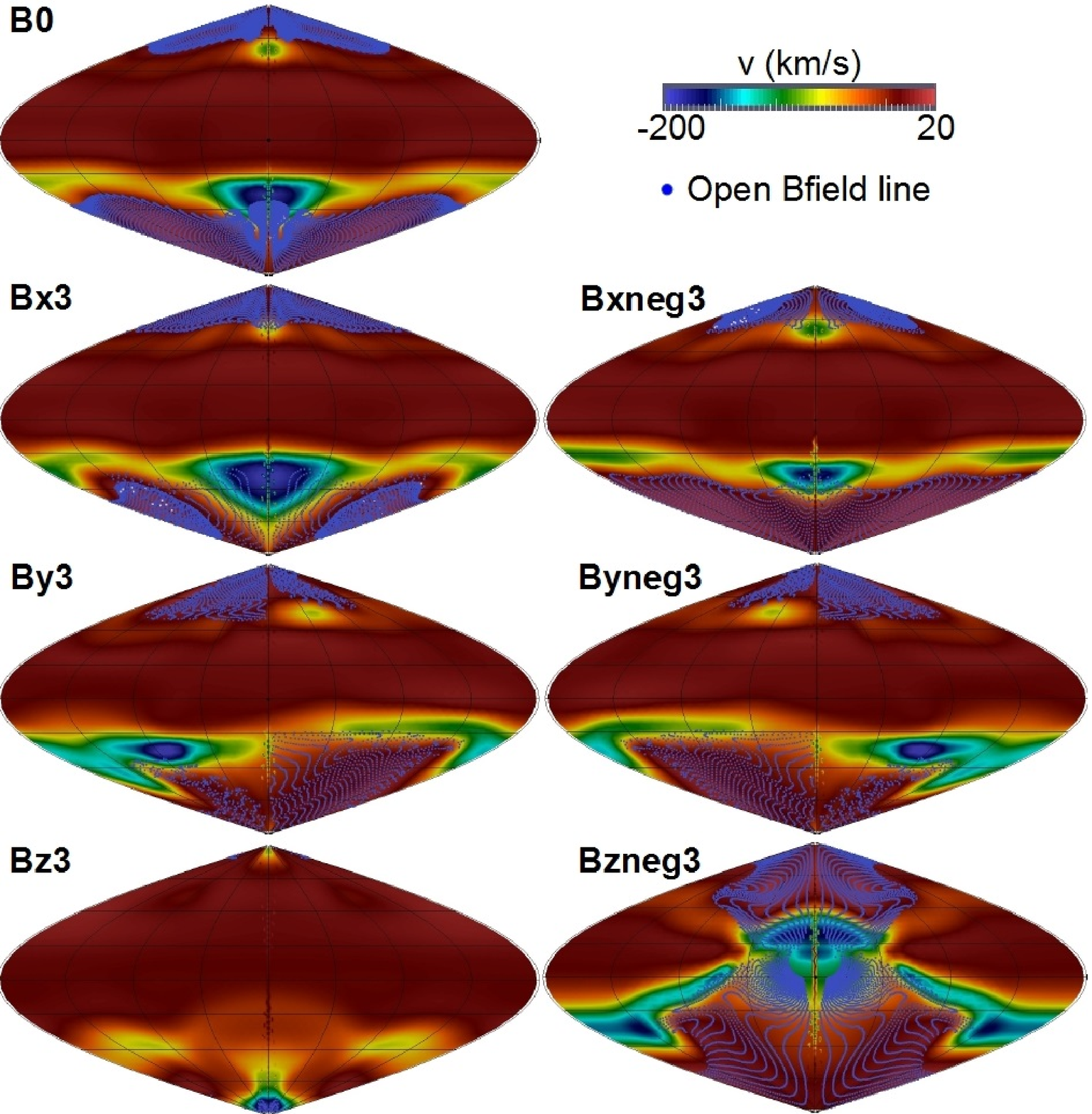}
\caption{Sinusoidal (Sanson-Flamsteed) projection of the radial velocity and open magnetic field lines (blue dots) on the Hermean surface (same simulations than Fig. 2).}
\end{figure}

Table 3 shows the integrated mass deposition at the North and South Hemisphere. Figure 10 shows the mass deposition distribution in the reference case (G-N, G-S), simulations Bx3 (A-N, A-S), Bxneg3 (D-N, D-S), By3 (B-N, B-S), Byneg3 (E-N, E-S), Bz3 (C-N, C-S) and Bzneg2 (F-N, F-S).

\begin{table}[h]
\centering
\begin{tabular}{c | c c}
Model & North Hemisphere & South Hemisphere  \\ \hline
Reference & $0.0305$ & $0.0679$ \\
Bx3 & $0.0263$ & $0.1191$ \\
Bxneg3 & $0.0631$ & $0.0433$ \\
By3 & $0.0245$ & $0.0566$ \\
Byneg3 & $0.0256$ & $0.0578$ \\
Bz3 & $0.0190$ & $0.0412$ \\
Bzneg2 & $0.1272$ & $0.1006$ \\
\end{tabular}
\caption{Integrated mass deposition at the North and South Hemisphere (kg/s) in the reference case and same simulations than Fig. 2 (case Bzneg2 instead of simulation Bzneg3).}
\end{table}

The regions with mass deposition in Bx3 (Bxneg3) simulation are narrower (wider) at the North Hemisphere and wider (narrower) at the South Hemisphere compared to the reference case. There is an East-West asymmetry of the mass deposition distributions in By3 and Byneg simulations correlated to the asymmetry of the magnetosphere. Mass deposition regions at the North and South Hemispheres are smaller and located closer to the poles in Bz3 simulation. The deposition regions in Bzneg3 simulation are moved to the equator, wider compared to the reference case. The largest integrated mass deposition is observed in Bzneg3 simulation and the weakest in Bz3 case, $232 \%$ and $61 \%$ respectively compared to the reference case. The mass deposition drops at the North (South) Hemisphere a $86 \%$ ($64 \%$) and increases at the South (North) Hemisphere a $175 \%$ ($207 \%$) in Bx3 (Bxneg3) simulation compared to the reference case. The magnetosphere East-West asymmetry leads to a decrease of the integrated mass deposition at both Hemispheres ($82 \%$ at the North and $84 \%$ at the South Hemisphere).

\begin{figure}[h]
\centering
\includegraphics[width=0.8\textwidth]{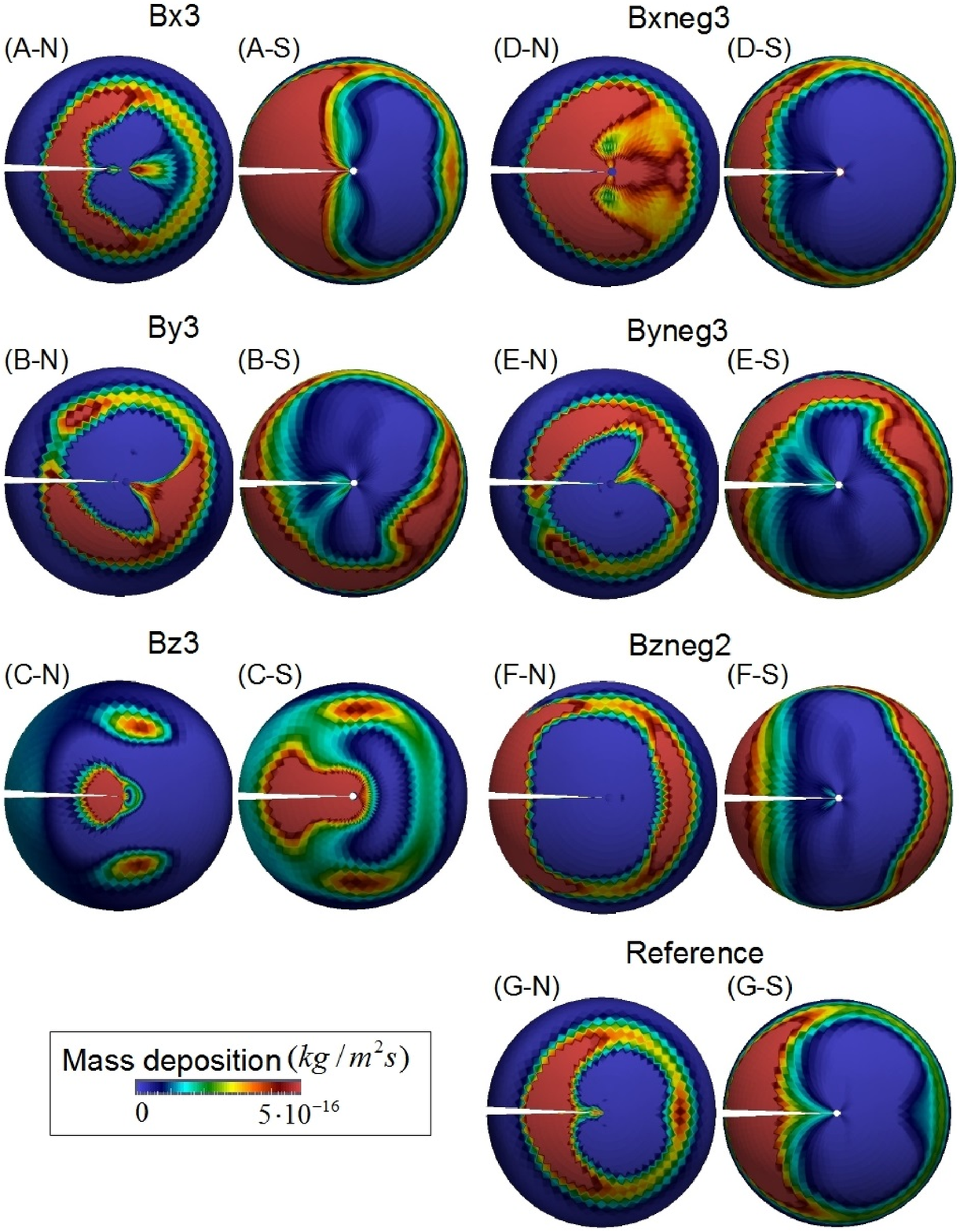}
\caption{Mass deposition at the North and South Hemisphere in the reference case (G-N, G-S), simulation Bx3 (A-N, A-S), Bxneg3 (D-N, D-S),By3 (B-N, B-S), Byneg3 (E-N, E-S), Bz3 (C-N, C-S) and Bzneg2 (F-N, F-S).}
\end{figure}

Figure 11 shows the energy deposition on the Hermean surface ($E = m_{p}v^{2}/2$) and regions of efficient particle sputtering. We define a region of efficient sputtering if the energy of the arriving particles is $E \geq 2$ eV, the minimum energy to overcome the surface binding energy of several chemical elements on the Hermean surface \citep{Kudriavtsev2005273}. A large extent of the energy deposition in the reference case takes place at the South Hemisphere day side. At the North Hemisphere there is only a small region of efficient particle sputtering nearby the pole. Bx3 (Bxneg3) simulation shows an enhancement of the energy deposition and regions of effective particle sputtering at the South (North) Hemisphere, due to the presence of a reconnection region in the South (North) of the magnetosphere. In By3 (Byneg3) simulation, the energy deposition is smaller at both Hemispheres. The local maxima of energy deposition and efficient sputtering are moved to the West (East) at the South Hemisphere and to the East (West) at the North Hemisphere. In Bz3 (Bzneg2) simulation the maxima of the energy deposition is smaller (larger) and the regions of effective sputtering are narrower (wider), located closer to the poles (equator). 

\begin{figure}[h]
\centering
\includegraphics[width=0.8\textwidth]{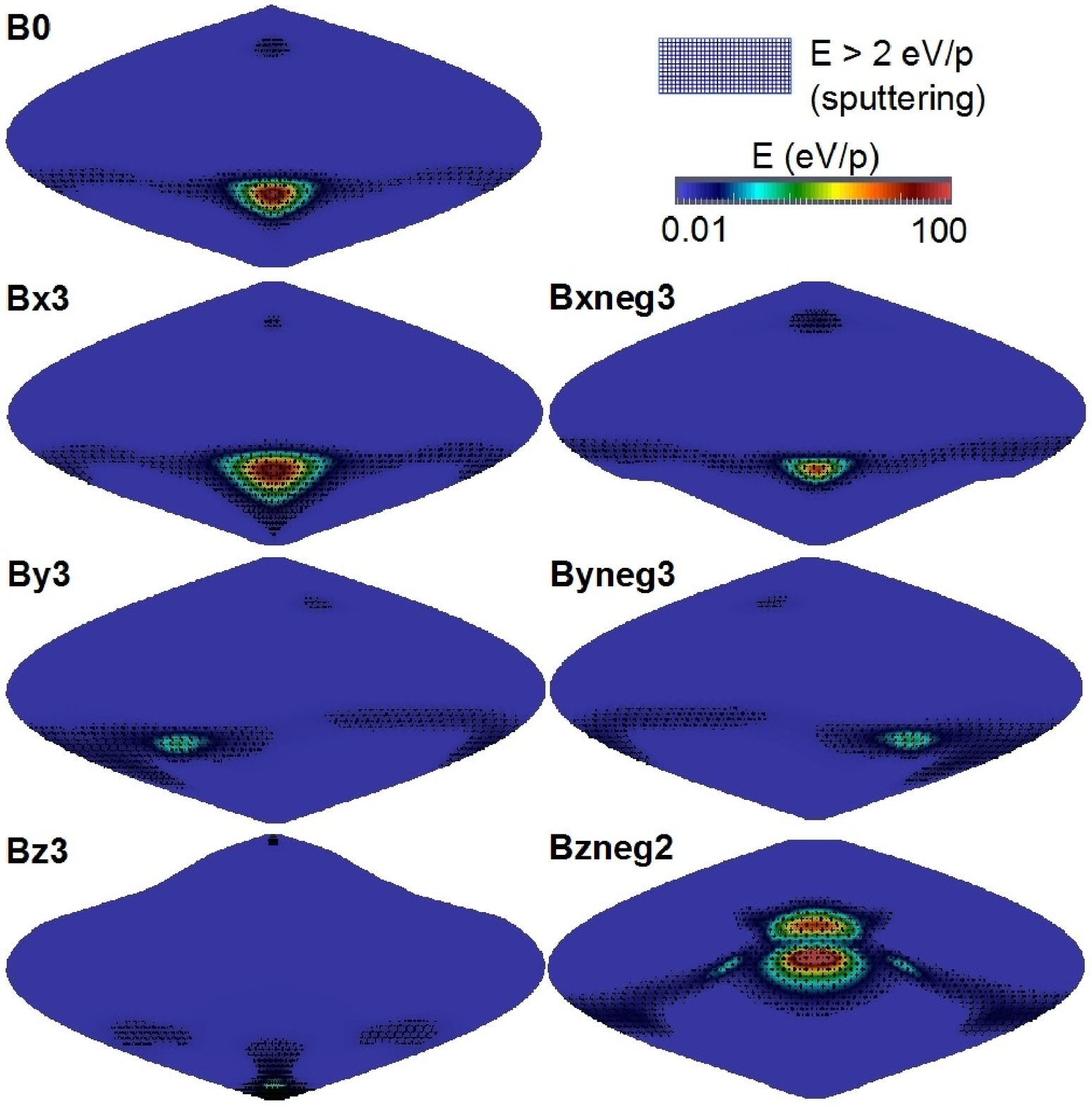}
\caption{Sinusoidal (Sanson-Flamsteed) projection of the energy deposition on the Hermean surface ($E = m_{p}v^{2}/2$) and regions of efficient particle sputtering ($E \geq 2$ eV). Same simulations than Fig. 2, except case Bzneg2 instead of simulation Bzneg3).}
\end{figure}

Figure 12 shows the averaged mass deposition (A) and the area of effective sputtering on the Hermean surface (B). The mass deposition for a Bx (Bxneg) IMF orientation decreases (increases) at the North Hemisphere if the IMF intensity is enhanced, although it increases (decreases) at the South Hemisphere. In Bx3 (Bxneg3) simulation the mass deposition decreases a $15 \% $ (2 times larger) at the North Hemisphere and increases a $60 \%$ (decreases a $65 \%$) at the South Hemisphere compared to the reference case. For a By-Byneg orientation the mass deposition at the North Hemisphere is almost the same in all simulations, but it decreases at the South Hemisphere if the IMF intensity increases, declining a $50 \%$ in By3 and Byneg3 simulations compared to the reference case. For a Bz (Bzneg) orientation the mass deposition decreases (increases) at both Hemisphere if the IMF intensity increases. In Bz3 (Bzneg2) simulation the mass deposition decreases a $65 \%$ (increases almost 4 times) at the North Hemisphere and a $65 \%$ (more than 2 times) at the South Hemisphere.

The area of efficient sputtering increases for a Bx IMF orientation if the IMF intensity increases. For a Bxneg IMF orientation the exposed area only grows slightly if the IMF intensity increases, because the reconnection region in the North of the magnetosphere leads to a reduction of the particle sputtering at the South Hemisphere, that almost compensates the enhancement at the North Hemisphere. For a By-Byneg orientation the sputtering area only increases slightly with the IMF intensity, a $18 \%$ for the simulations By3 and Byneg3 compared to the reference case. For a Bz (Bzneg) orientation the area decreases (increases) with the IMF intensity. In Bz3 simulation the sputtering area decreases by half of the reference case and it is almost three times bigger in Bzneg2 simulation.

\begin{figure}[h]
\centering
\includegraphics[width=0.8\textwidth]{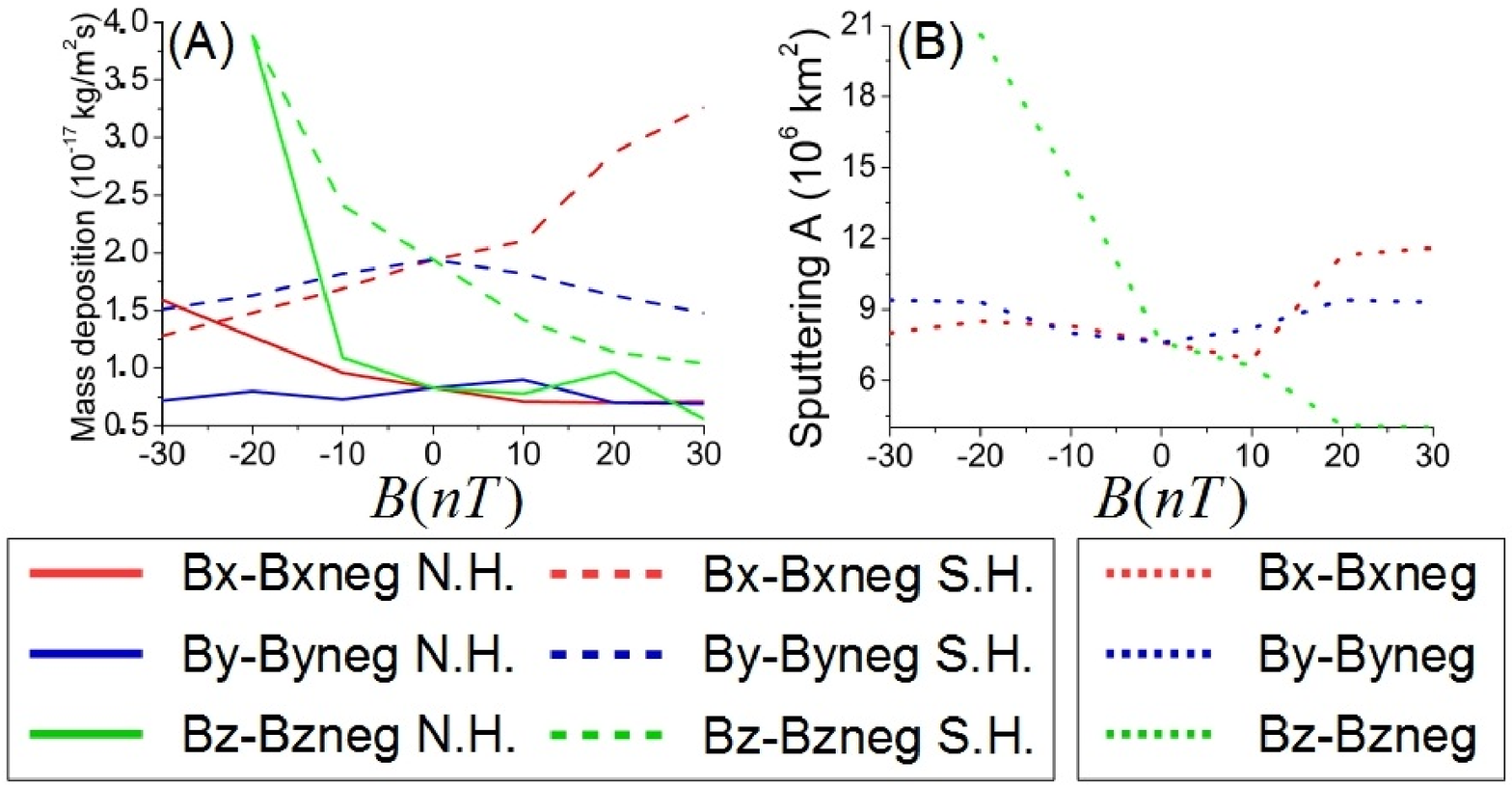}
\caption{Averaged mass deposition (A) and area of effective sputtering on the Hermean surface (B) for different IMF orientations and intensities.}
\end{figure}

\section{Conclusions and discussion}
\label{Conclusions}

The different IMF configurations under study induce specific distortions in the topology of the Hermean magnetic field, altering the location of the reconnection regions as well as the particle precipitation and sputtering on the planet surface. For a Southward IMF orientations (Bzneg model) the erosion driven by the IMF on the day side of the Hermean magnetic field leads to the direct precipitation of the SW at the planet equatorial region if the IMF module is larger than $20$ nT. The magnetopause stand off distance is almost independent of the IMF intensity for Sun-Mercury (Bxneg model) or Mercury-Sun (Bx model) IMF orientations, located further away from the planet if the IMF is Northward (Bz model), due to an enhancement of the Hermean magnetic field at the equatorial region. IMF orientations towards West (Byneg model) or East (By model) directions on the planet orbital plane induce a West-East asymmetry in the Hermean magnetosphere, leading to a decrease of the stand of distance.

Bx-Bxneg IMF orientation leads to the magnetosphere configuration with the most compressed magnetosheath at low latitudes, compression enhanced as the IMF intensity increases, although the compression is weaker at high latitudes. The magnetosheath for a By-Byneg IMF orientation is less compressed as the IMF intensity increases and the magnetopause is slightly closer to the planet. The magnetosheath is less compressed at low latitudes and more compressed at high latitudes for a Bz IMF orientation, with the magnetopause located further away from the planet as the IMF intensity increases (opposite scenario for a Bzneg IMF orientation).

The plasma is injected into the inner magnetosphere through the reconnection regions, leading to the formation of a plasma stream that connects magnetosheath and planet surface, correlated with a local maxima of the mass deposition and particle sputtering. A large proportion of the mass deposition and particle sputtering takes place at the South Hemisphere day side, but there is a strong dependence with the IMF intensity and orientation at both Hemispheres, moving and modifying the magnitude of the local maxima. Compared to the reference case, for a Bx IMF orientation there is an enhancement of the mass deposition and sputtering at the South Hemisphere and a weakening at the North Hemisphere, due to the presence of a wide reconnection region in the South of the magnetosphere. A Bxneg orientation leads to the opposite scenario because the reconnection region is in the North of the magnetosphere. If the IMF intensity increases, the mass deposition and the area of efficient sputtering increase more for a Bx than for a Bxneg IMF orientation, $30 \%$ higher mass deposition and $50 \%$ stronger particle sputtering, because for a Bzneg IMF orientation the increase at the North Hemisphere is almost compensated by the reduction at the South Hemisphere. The East-West asymmetry of the magnetosphere driven by a By-Byneg IMF orientation leads to East-West displacements of the reconnection regions, and a reduction of the mass deposition and particle sputtering. Mass deposition decreases if IMF intensity increases although particle sputtering slightly increases, because the West-East asymmetry weakness the plasma stream, more spread when it reaches the surface of Mercury. For a Bz IMF orientation the local maxima of the mass deposition and particle sputtering decrease, moved to the poles, although for a Bzneg IMF orientation both maxima increase and are displaced at the planet equator. Mass deposition and the area of efficient sputtering are enhanced for a Bzneg IMF orientation and weakened for a Bz IMF orientation if the IMF intensity increases. Mass deposition is almost 8 times larger and particle sputtering is 10 times larger if we compare Bzneg2 and Bz3 simulations. Bzneg2 simulation shows the largest integrated mass deposition, but the largest integrated mass deposition at the South Hemisphere is observed in Bx3 simulation.

The area covered by open magnetic field lines increases with the IMF intensity for Bx-Bxneg and By-Byneg IMF orientations, 3 times larger at the South Hemisphere than at the North Hemisphere. The exposed area decreases for a Bz IMF orientation and increases for a Bzneg IMF orientation. The ratio of area covered by open field lines at the North and South Hemisphere remains almost constant for Bz IMF orientations if the IMF intensity increases. For a Bzneg IMF orientation the ratio changes between 3 times larger at the South Hemisphere if the IMF intensity is $10$ nT to a $80 \%$ if the IMF intensity is $30$ nT. 

The oval angle at the North Hemisphere is $30^{o}$ for a Bx-Bxneg IMF orientation and $50^{o}$ at the South Hemisphere, weakly associated with the IMF intensity, only $5^{o}$ larger comparing the reference case to the simulation of IMF intensity $30$ nT. A By-Byneg IMF orientation shows different oval angles in West and East sides of the Hemispheres, due to the West-East asymmetry of the magnetosphere, as well as a stronger dependency with the IMF intensity, increasing from $30^{o}$ to $40^{o}$ at the North Hemisphere and from $50^{o}$ to $70^{o}$ at the South Hemisphere compared to the reference case. A Bz-Bzneg IMF orientation leads to an asymmetry of the oval angle on the day and night side of the planet, strongly dependent with the IMF intensity. The oval angle on the day side reaches the equator at both Hemispheres in Bzneg3 simulation, although it is $9^{o}$ at the North and $14^{o}$ at the South Hemisphere in Bz3 simulation.

Observational studies of protons precipitation during a Southward IMF of 10 nT found a mean open magnetic field area of $2.8 \cdot 10^{6}$ km$^{2}$ and a mean proton flux of $4.1 \cdot 10^{18}$ km$^{-2}$ s$^{-1}$ \citep{2003Icar..166..229M}, flowing down into the cusp from the magnetosheath and lost to surface precipitation \citep{2014JGRA..119.6587R}, results compatible with present study. Temporal and latitudinal variability of charged particles precipitation is also reproduced in models with different IMF orientation and intensity \citep{2012LPI....43.1646D}.
Periods of stable northward IMF without plasma depletion are identified in the simulation set as the models of lowest plasma precipitation on the planet surface \citep{2013JGRA..118.7181G}, Periods of southward IMF, characterized by large flux transfer events in the magnetosheath, are identified as the models of largest plasma precipitation, intense magnetosheath plasma depletion and the strongest particle sputtering on the planet surface \citep{2013AGUFMSM24A..03D,2007SSRv..132..433K}. The plasma depletion layer is not resolved as a decoupled global structure from the magnetosheath due to a lack of model resolution, although simulations and observations share similar features in between the magnetosheath and magnetopause. The magnetic diffusion of the model is several orders of magnitude larger than the real plasma, so the reconnection between interplanetary and Hermean magnetic field is instantaneous (no magnetic pile-up on the planet dayside) and stronger (enhanced erosion of the magnetic field of Mercury), although the essential role of the reconnection region in the depletion of the magnetosheath and the injection of plasma into the inner magnetosphere is reproduced.

\section*{Aknowledgments}
The research leading to these results has received funding from the European Commission's Seventh Framework Programme (FP7/2007-2013) under the grant agreement SHOCK (project number 284515). The MESSENGER magnetometer data set was obtained from the NASA Planetary Data System (PDS) and the values of the solar wind hydrodynamic parameters from the NASA Integrated Space Weather Analysis System.

\section*{Appendix}
Summary of simulations parameters:

\begin{table}[h]
\centering
\begin{tabular}{c | c c c c}
Model & $\vec{B}$ (nT) & $n$ (cm$^{-3}$) & $T$ ($10^{5}$ K) & $\mathrm{v}$ (km/s) \\ \hline
Reference & (0, 0, 0) & $60$ & 0.58 & 250 \\
Bx & (10, 0, 0) & $60$ & 0.58 & 250 \\
Bx2 & (20, 0, 0) & $60$ & 0.58 & 250 \\
Bx3 & (30, 0, 0) & $60$ & 0.58 & 250 \\
Bxneg & (-10, 0, 0) & $60$ & 0.58 & 250 \\
Bxneg2 & (-20, 0, 0) & $60$ & 0.58 & 250 \\
Bxneg3 & (-30, 0, 0) & $60$ & 0.58 & 250 \\
By & (0, 10, 0) & $60$ & 0.58 & 250 \\
By2 & (0, 20, 0) & $60$ & 0.58 & 250 \\
By3 & (0, 30, 0) & $60$ & 0.58 & 250 \\
Byneg & (0, -10, 0) & $60$ & 0.58 & 250 \\
Byneg2 & (0, -20, 0) & $60$ & 0.58 & 250 \\
Byneg3 & (0, -30, 0) & $60$ & 0.58 & 250 \\
Bz & (0, 0, 10) & $60$ & 0.58 & 250 \\
Bz2 & (0, 0, 20) & $60$ & 0.58 & 250 \\
Bz3 & (0, 0, 30) & $60$ & 0.58 & 250 \\
Bzneg & (0, 0, -10) & $60$ & 0.58 & 250 \\
Bzneg2 & (0, 0, -20) & $60$ & 0.58 & 250 \\
Bzneg3 & (0, 0, -30) & $60$ & 0.58 & 250 \\

\end{tabular}
\caption{Summary of simulations parameters.}
\end{table}
 
\section*{References}

\bibliography{mybibfile}

\end{document}